\documentclass[12pt,preprint]{aastex}

\shorttitle{Classical Cepheids trace an inner thin disk}
\shortauthors{D\'ek\'any et al.}

\begin{document}

\title{The VVV Survey reveals classical Cepheids tracing a young and thin stellar disk across the Galaxy's bulge}

\author{I.~D\'ek\'any\altaffilmark{1,2},
D.~Minniti\altaffilmark{3,1,4},
D.~Majaess\altaffilmark{5,6},
M.~Zoccali\altaffilmark{2,1},
G.~Hajdu\altaffilmark{2,1},
J.~Alonso-Garc\'ia\altaffilmark{7,1},
M.~Catelan\altaffilmark{2,1},
W.~Gieren\altaffilmark{8,1},
J.~Borissova\altaffilmark{9,1}}

\affil{$^1$Instituto Milenio de Astrof\'isica, Santiago, Chile.}
\affil{$^2$Instituto de Astrof\'isica, Facultad de F\'isica, Pontificia Universidad Cat\'olica de Chile, Av. Vicu\~na Mackenna 4860, Santiago, Chile.}
\affil{$^3$Departamento de F\'isica, Facultad de Ciencias Exactas, Universidad Andres Bello, Rep\'ublica 220, Santiago, Chile.}
\affil{$^4$Vatican Observatory, V00120 Vatican City State, Italy.}
\affil{$^5$Saint Mary's University, Halifax, Nova Scotia, Canada.}
\affil{$^6$Mount Saint Vincent University, Halifax, Nova Scotia, Canada.}
\affil{$^7$Unidad de Astronom\'ia, Universidad de Antofagasta, Av. U. de Antofagasta 02800, Antofagasta, Chile.}
\affil{$^8$Departamento de Astronom\'ia, Universidad de Concepci\'on, Casilla 160-C, Concepci\'on, Chile}
\affil{$^9$Instituto de F\'isica y Astronom\'ia, Universidad de Valpara\'iso, Av. Gran Breta\~na 1111, Valparaíso, Chile.}

\begin{abstract}
Solid insight into the physics of the inner Milky Way is key to understanding our Galaxy's evolution, but extreme dust obscuration has historically hindered efforts to map the area along the Galactic mid-plane. New comprehensive near-infrared time-series photometry from the VVV Survey has revealed 35 classical Cepheids, tracing a previously unobserved component of the inner Galaxy, namely a ubiquitous inner thin disk of young stars along the Galactic mid-plane, traversing across the bulge. The discovered period (age) spread of these classical Cepheids implies a continuous supply of newly formed stars in the central region of the Galaxy over the last $100$ million years.
\end{abstract}


\section{Introduction}
The inner Milky Way is dominated by a peanut-shaped bulge \citep{2010ApJ...724.1491M,2013MNRAS.435.1874W} that flares up from a prominent Galactic bar \citep{1991Natur.353..140N}, and its structural and kinematical properties are consistent with a formation scenario driven by the instabilities of a multi-component stellar disk \citep{2013MNRAS.432.2092N,2014A&A...562A..66Z,2014MNRAS.438.3275G,2014ApJ...787L..19N,2015A&A...577A...1D}.
The majority of its constituent stars are very ($\gtrsim 8$ billion years) old \citep{1995Natur.377..701O,2003A&A...399..931Z,2010ApJ...725L..19B}, implying that its formation occurred at early epochs of the Milky Way's evolution.
Although intermediate-age stars are also present in the bulge \citep{2003MNRAS.338..857V,2013A&A...549A.147B}, their origin, nature and ubiquity are poorly understood, owing partly to sizable age uncertainties and possible biases from small sample sizes and contamination by foreground disk stars.
Yet it is well known that the innermost core ($R\lesssim300~{\rm pc}$) of the Galaxy hosts stars with ages ranging from a few million to several billion years \citep[e.g.,][]{1996Natur.382..602S,2003MNRAS.338..857V,2011Natur.477..188M}. This ``nuclear bulge'' has rather distinct physical properties from the rest of the bulge and is related to the Central Molecular Zone, but neither its transition to the surrounding bulge regions nor the triggering process for star formation and its history are well understood \citep[see, e.g.,][]{2002A&A...384..112L}.

Our global picture of the inner Galaxy is linked primarily to observations of the outer bulge, i.e., at Galactic latitudes higher than $\sim2^{\circ}$. That deficit arises from extreme obscuration by interstellar dust, high source density, and confusion with foreground disk populations. The properties of the bulge region at low latitudes have thus remained largely unexplored, leaving ambiguities concerning the interplay of the nuclear bulge and the various components of the boxy bulge and surrounding outer stellar disk. The VISTA Variables in the V\'ia L\'actea (VVV) ESO Public Survey \citep{2010NewA...15..433M} presents a means to ameliorate the situation 
by opening the time-domain in the near-infrared. That is particularly important as classes of variable stars, such as classical Cepheids, are indicators of young stellar populations \citep[e.g.,][]{2015pust.book.....C}, and can yield critical insight as demonstrated here.

\section{Discovery and Characterisation of the Cepheids}

We performed a comprehensive near-infrared variability search using VVV Survey data collected between 2010 and 2014, in a $\sim66$ square degree area in the central bulge ($-10.5^{\circ}\lesssim l \lesssim +10^{\circ}, -1.7^{\circ} \lesssim b \lesssim +2^{\circ}$, aligned with VVV image borders). We excluded two fields lying toward the nuclear bulge owing to extreme source crowding. Time-series photometry for $\sim10^8$ point sources were analyzed, with up to 70 $K_s$-band measurements per object, together with color information from 1--5 independent measurements in the $Z$, $Y$, $J$, and $H$ bands. The data processing, calibration, and light-curve analysis were conducted in the same way as in our previous study on the Twin Cepheids beyond the bulge \citep{2015ApJ...799L..11D}, and are based on standard VVV Survey data products \citep{2010NewA...15..433M,2004SPIE.5493..411I,2013arXiv1310.1996C}. We detected a sample of approximately $3\cdot10^5$ objects which displayed putative light variations, and scanned the results for Cepheids possessing pulsation periods in the range of 4--50 days. The lower limit mitigates confusion between the pulsation modes \citep[e.g.,][]{2015AJ....149..117M}, while the upper limit was constrained by the sampling of the photometric time-series. We found 655 fundamental-mode Cepheid candidates based on their periods, amplitudes, and asymmetric light-curve shapes. A large fraction of these objects were only detected in the $H$ and $K_s$ bands given the extreme reddening.

For computing the distances and extinctions of the Cepheids, we employed the period-luminosity (PL) relations \citep[a.k.a. Leavitt Law,][]{2012ApJ...759..146M,2009MNRAS.397..933M} adopted in our previous study \citep{2015ApJ...799L..11D}, adjusted to match the latest and most accurate distance modulus for the Large Magellanic Cloud \citep[LMC,][]{2013Natur.495...76P}. The color excess was converted to an extinction using a selective-to-absolute extinction ratio found towards the Galactic Center \citep[GC,][]{2009ApJ...696.1407N}, once converted into the VISTA photometric system \citep[see][]{2015ApJ...799L..11D}. For each object, a distance and extinction were computed under both the assumptions that the target is a classical and a type II Cepheid. We subsequently identified the correct solution and class upon further analysis.

Following the approach adopted in our previous study \citep{2015ApJ...799L..11D}, uncertainties tied to the distance and extinction were estimated by Monte Carlo simulations.  The distances exhibit 1-3\% precision and 8-10\% accuracy. The latter is dominated by the uncertainty in the adopted extinction curve. The systematic uncertainty might be larger, possibly up to $\sim20\%$, if variations in the extinction's wavelength dependence, either as a function of Galactic longitude or along the sight-line, significantly exceed those found previously toward the inner bulge \citep{2009ApJ...696.1407N}.

Type II Cepheids, i.e. old, low-mass, He-shell burning pulsating stars in the classical instability strip \citep{2015pust.book.....C}, populate the same range of periods as classical Cepheids.  The two Cepheid classes share similar near-infrared light-curves \citep[see, e.g.][]{2015pust.book.....C,2011Natur.477..188M,2015ApJ...799L..11D}. Since the bulge contains a sizable population of type II Cepheids with a centrally concentrated distribution \citep{2011AcA....61..285S}, our sample is expected to be dominated by these objects. Therefore, additional observational information is required to identify classical Cepheids in our sample. The critical information is conveyed by an extinction map of the bulge, based on the analysis of its red clump stars \citep{2012A&A...543A..13G}.

\begin{figure}
\epsscale{1.0}
\plotone{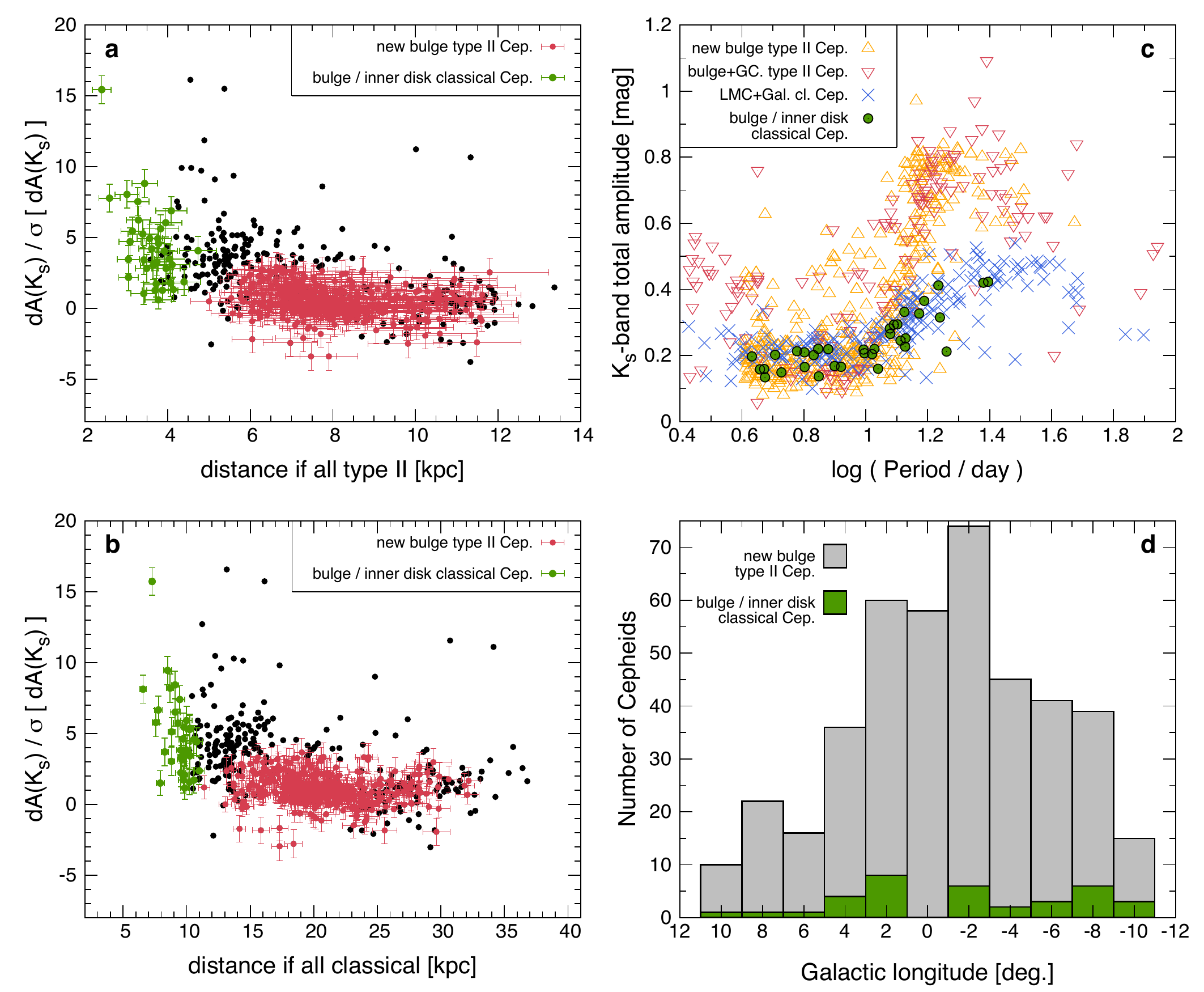}
\caption{The selection of classical Cepheids.
{\bf Panel a:} Under the assumption that {\em all Cepheids} in our sample are of type II, this figure shows the deviation of their computed extinction relative to the VVV reddening map \citep{2012A&A...543A..13G} divided by the uncertainty, as a function of the distance. Error bars show $1\sigma$ uncertainty ranges. \emph{Green points:} the 35 classical Cepheids inside the bulge volume; \emph{red points:} new type II bulge Cepheids with solid classifications; \emph{black points:} rest of the sample. Black points at short distances in this figure represent candidate classical Cepheids beyond the bulge, while those at larger distances are type II Cepheid candidates outside the bulge or objects with uncertain classification.
{\bf Panel b:} As panel a, but showing values computed under the assumption that {\em all Cepheids} are of the classical type.
{\bf Panel c:} $K_s$ total amplitude \emph{vs} $\log{P}$ diagram of classical Cepheids from the LMC \citep{2004AJ....128.2239P} and the Galactic field \citep{2011ApJS..193...12M} \emph{(blue crosses)}, previously known type II Cepheids in the bulge \citep[][$K_s$ amplitudes derived from VVV data]{2011AcA....61..285S} and in Galactic globular clusters \citep{2006MNRAS.370.1979M} \emph{(red triangles)}, and new type II Cepheids from our study \emph{(orange triangles)}. \emph{Green points} show the new classical Cepheids in the inner Galaxy.
\emph{\bf Panel d:} Histograms of the discovered classical and type II Cepheid sample as a function of Galactic longitude.\label{type}
}
\end{figure}

Classical and type II Cepheids have similar intrinsic colors but rather different luminosities \citep{2012ApJ...759..146M,2009MNRAS.397..933M}, and consequently their PL relations yield similar extinctions but very different distances when applied to the same object. A Cepheid's type can be determined if its distance and extinction are inconsistent with the cumulative extinction up to the bulge in its direction, computed under one of the two possible assumptions for its type \citep{2011Natur.477..188M,2015ApJ...799L..11D}. Figure.~\ref{type}a-b show the difference between the extinction values predicted by the PL relations and the bulge extinction map, normalized by its uncertainty as a function of distance, when extinction and distance are computed under the assumption that \emph{all} detected Cepheids are either classical or type II. There are several Cepheids in our sample for which the computed extinction was within $3\sigma$ agreement with, or higher than the predicted cumulative extinction up to the bulge towards their sight-lines, but for which the type II assumption yields short distances ($3-4$ kpc), making them \emph{bona fide} classical Cepheid candidates. While most of them reside in the disk at the far side of the bulge as expected, we found 35 stars (green points in Fig.~\ref{type}) that are likewise located within the bulge volume \citep[see, e.g.,][]{2013MNRAS.435.1874W}.

\begin{figure}
\epsscale{1.0}
\plotone{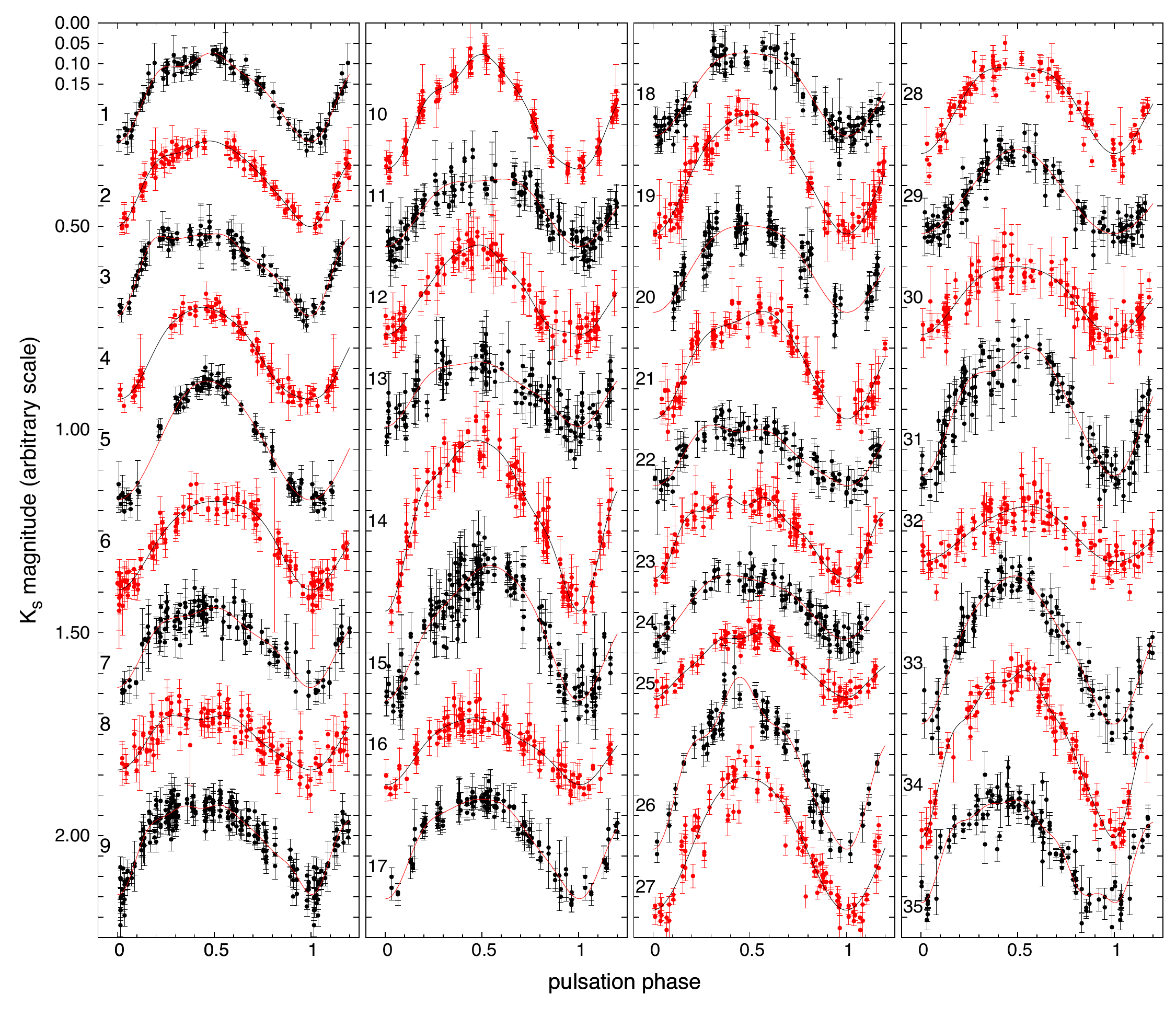}
\caption{$K_s$-band phase diagrams of the classical Cepheids in the inner Galaxy (on arbitrary magnitude scale). Curves denote Fourier series fits to the data, identifiers are shown on the left.\label{lc}}
\end{figure}
\begin{figure}
\epsscale{0.8}
\plotone{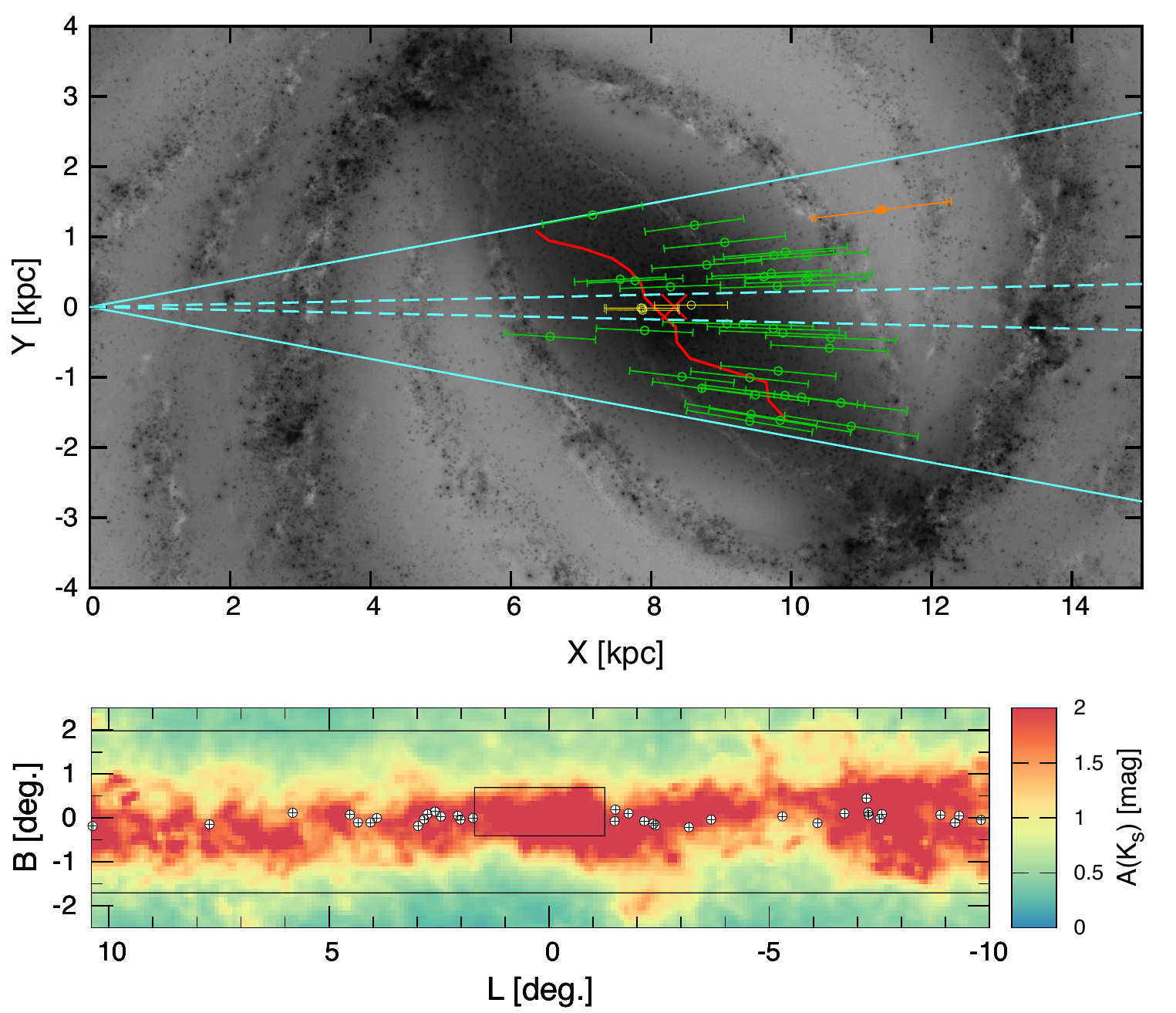}
\caption{\emph{Top panel}: positions of the classical Cepheids projected onto the Galactic plane and their errors \emph{(green symbols)}, overlaid on Robert Hurt's illustration of the Milky Way. \emph{Orange symbols} mark the Twin Cepheids \citep{2015ApJ...799L..11D} beyond the bulge, \emph{yellow symbols} represent the classical Cepheids in the nuclear bulge \citep{2011Natur.477..188M,2015ApJ...799...46M}. The Galactic bar \citep{2012A&A...543A..13G} is outlined by a red curve, whereas the \emph{red ``x''} marks the GC \citep{2013ApJ...776L..19D}. The \emph{solid and dashed cyan lines} denote the longitudinal limits of our study and the direction of the gap in coverage towards the GC, respectively. The \emph{bottom panel} shows the classical Cepheids' positions in Galactic coordinates, overlaid on the VVV bulge $K_s$ extinction map \citep{2012A&A...543A..13G}. \emph{Black lines} denote the borders of the surveyed area.\label{map}}
\end{figure}

These 35 stars could be type II Cepheids only if they were located by chance beyond thick dust clouds of small angular sizes ($<1'$), otherwise the $2'\times2'$ resolution of the VVV reddening map would be sensitive to those. To investigate this unlikely possibility, we examined the color images and analyzed the color-magnitude diagrams (CMD's) of the stellar fields around these Cepheids, in search of anomalously high gradients in the foreground extinction. In all cases, the CMDs are inconsistent with the presence of such nearby foreground clouds, down to angular sizes of $\sim10''$, where the surface density of the disk population becomes too low for drawing such conclusion. That limiting angular scale would correspond to physical sizes of $\lesssim0.2$~pc, smaller than a typical Bok globule \citep{2015MNRAS.452..389D}. The probability for such chance alignment is negligible, and that it affects all 35 objects is virtually zero. Consequently, we rule out the possibility that they are nearby type II Cepheids in the Galactic disk, and classify them as classical Cepheids located inside the bulge volume. Their light-curves are presented in Fig.~\ref{lc}, while Fig.~\ref{map} conveys their spatial distribution. Their properties are summarized in Table~\ref{table}.

By reversing the arguments above, 425 new type II Cepheids are identified in the central bulge (Fig.~\ref{type}a-b, red points), with extinctions consistent with predictions from the VVV reddening map \citep{2012A&A...543A..13G}. If these stars were classical Cepheids, they would be located far beyond the bulge, and their corresponding extinctions would be inconsistent with such large distances. The detailed study of these objects and the rest of the new classical Cepheids beyond the bulge will be presented in forthcoming papers.

To bolster the classification of the classical and type II Cepheids in our sample, the distributions of their amplitudes and positions were compared. The middle panel of Fig.~\ref{type} displays the period-amplitude diagram of these objects with previously known classical and type II Cepheids. Although these parameters do not allow us to unambiguously separate the two types, particularly at shorter periods, the locus of the newly discovered classical Cepheid population agrees with that of other classical Cepheids in the Local Group. Furthermore, the right panel of Fig.~\ref{type} compares the spatial distributions of classical and type II Cepheids identified in this study. While the type II objects are expectedly concentrated towards the GC \citep{2011AcA....61..285S}, the classical Cepheids are rather evenly distributed. The lack of classical Cepheids toward small longitudes, and the slightly offset peak from zero longitude in the density of type II Cepheids, arise from the central gap in the coverage ({\em cf.} Fig.~\ref{map}).

The sample of 35 new classical Cepheids is incomplete owing to the bright magnitude limit of the VVV Survey. Brighter longer-period Cepheids \citep{2012ApJ...759..146M} can be saturated in the VVV images in the absence of significant foreground extinction. The asymmetry in the classical Cepheids' observed number density distribution, with more objects at the far side of the bulge (Fig.~\ref{map}), is an observational bias owing to that saturation limit. The typical $K_s$ limit where saturation cannot be calibrated out is $\sim11$~mag.  The faint detection limit towards the central bulge is $K_s\sim15.5$~mag. These estimates are time-varying, and depend on the seeing, sky transparency, and crowding. 

\begin{figure}
\epsscale{0.6}
\plotone{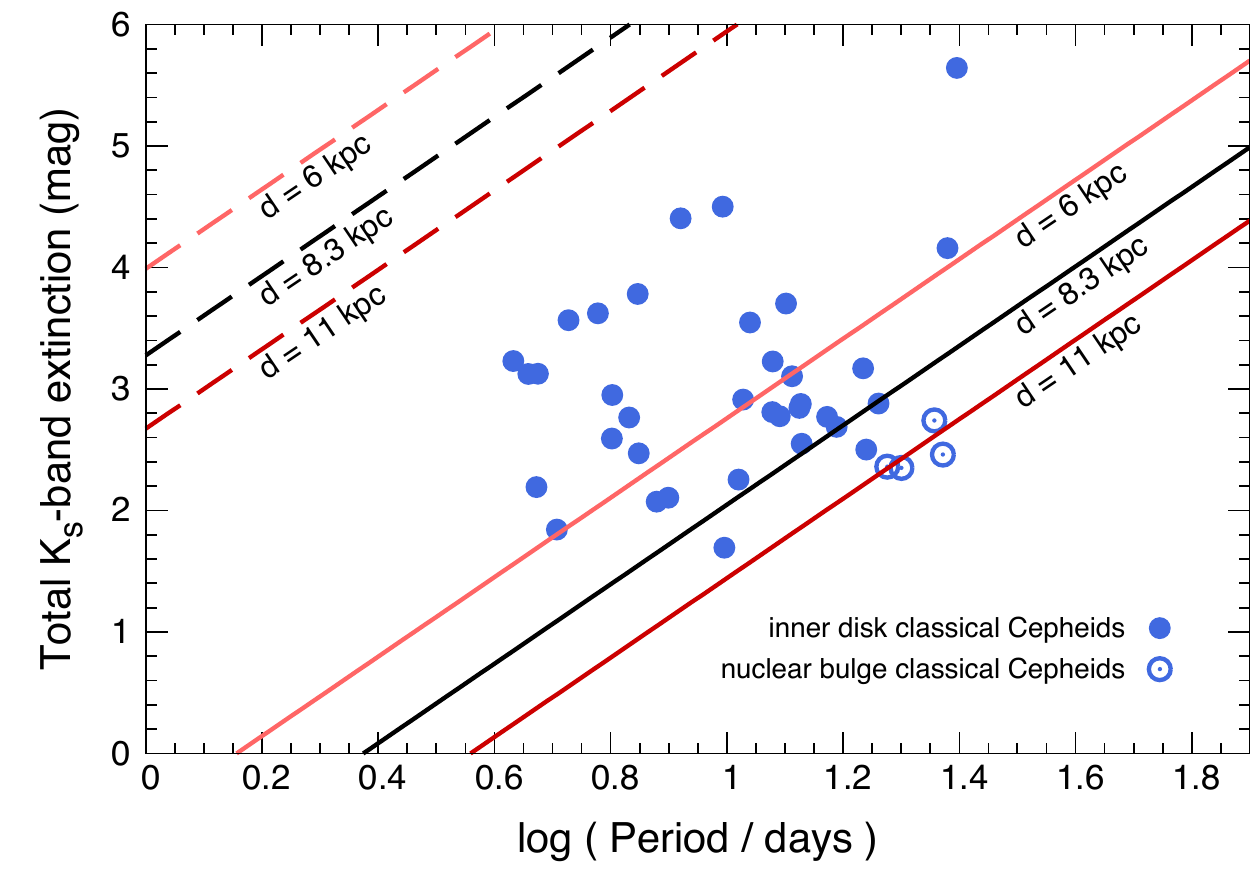}
\caption{Approximate VVV Survey detection limits, showing the range of pulsation periods and foreground $K_s$ extinction where classical Cepheids in the inner Galaxy can be detected. \emph{Dashed lines} delineate the faint detection limits, whereas \emph{solid lines} convey saturation limits for Cepheids at three different distances, illustrating the distance to the GC \citep{2013ApJ...776L..19D} and the corresponding near and far edges of the bulge \citep{2013MNRAS.435.1874W}. \emph{Blue points} mark the classical Cepheids discovered here, \emph{blue circles} show the values of classical Cepheids in the nuclear bulge \citep{2011Natur.477..188M,2015ApJ...799...46M}.\label{limits}
}
\end{figure}

Figure~\ref{limits} conveys the detection threshold in period--extinction parameter space for classical Cepheids, at various distances. Short-period objects near the $4^{\rm d}$ lower limit are detected throughout the entire range of observed extinction. However, nearly half of our sample would be overlooked in the absence of significant reddening if they were located at the near side of the bulge. Consequently, the asymmetry in the objects' observed spatial distribution, with more Cepheids further away (Fig.~\ref{map}), is an observational bias owing to the saturation limit of the VVV Survey.

On the other hand, the observed small vertical spread of the classical Cepheids cannot be attributed to a similar observational bias. While long-period classical Cepheids further from the Galactic plane and occupying the bulge would remain undetected by the VVV Survey, more than half our sample would still be detected if they were located at considerably higher Galactic latitudes. That is apparent when comparing the distributions of Fig.~\ref{limits} and the mean bulge extinction as a function of Galactic latitude in the VVV extinction map \citep[][see also Fig.~\ref{map}, lower panel]{2012A&A...543A..13G}. Bulge Cepheids exhibiting periods less than 15 days would be detected up to latitudes of $1^{\circ}$ at most longitudes, while those with periods below 8 days would be found at latitudes typically up to $2^{\circ}$.

\section{The inner thin disk}

The 35 new classical Cepheids inside the bulge volume span across the entire longitudinal range of our study and lie in close proximity to the Galactic mid-plane. The standard deviation of their vertical distances from it is only 22 pc, with the farthest Cepheid being only 82 pc above it. This vertical distribution is less than estimates of relatively nearby classical Cepheids \citep{2009MNRAS.398..263M}, whereby the latter exhibit a scale height of $\sim75$~pc.

All classical Cepheids discovered are younger than 100~Myr, since a Cepheid's pulsation period is closely linked to its age \citep{2005ApJ...621..966B}. The youngest Cepheid observed may be $\sim25$ million years old \citep{2005ApJ...621..966B}. We cannot exclude the possible presence of even younger and brighter Cepheids, which would be saturated in the VVV Survey. Our objects thus trace an underlying young and thin inner stellar disk along the Galactic plane. Although their census may be incomplete, it is noteworthy that both short- and long-period Cepheids were detected, while in the nuclear bulge only Cepheids with periods close to 20 days are known. The period spread implies that they originate from continuous star formation along the mid-plane in the central Galaxy over the last $\sim100$ million years. 

A limiting factor that hinders a detailed analysis of the spatial distribution of Cepheids in the inner Galaxy is the uncertainty in the wavelength dependence of interstellar extinction, i.e., the properties of interstellar dust particles 
along the sight-line. However, the bulk of the classical Cepheids in our sample remain within 3~kpc of the GC if we consider different extinction curves \citep[e.g.,][]{1989ApJ...345..245C,1999PASP..111...63F}, even if they have not been observed to be valid towards the inner bulge. For instance, adopting the ``standard'' Galactic extinction curve \citep{1989ApJ...345..245C} would shift the mean Cepheid distance $\sim1.5$~kpc closer, and only 2 objects would become further than 3 kpc from the GC (in the near disk). The discovery of a thin star-forming inner disk across the bulge is immune to the effects of potentially anomalous extinction.

Cepheid extinction estimates derived from PL relations were compared to the VVV extinction map \citep{2012A&A...543A..13G}. The comparison is nearly split between values that generally agree, and Cepheids that exhibit significant ($>3\sigma$) positive deviations from this map's predictions. These anomalies may originate from localized absorbing material rather than scale height differences between the two populations. Since bulge red clump stars are detected to small angular radii ($\sim10''$) around all the Cepheids, most of the material should be distributed inside the bulge's volume and along the sight-lines of objects with extreme extinction. Several of these Cepheids are in the angular vicinity ($\lesssim30''$) of ionized hydrogen bubbles, embedded young stellar objects, or extended mid-infrared objects visible in images of the GLIMPSE \citep{2009PASP..121..213C} surveys (see Table~\ref{table}).

\section{Conclusion}

The presence of numerous classical Cepheids demonstrate that the central Galaxy contains a very young ($<100$~Myr old) stellar population spanning well outside the nuclear bulge. Our results transcend earlier evidence for intermediate-age stars in the bulge volume \citep[e.g.,][]{2003MNRAS.338..857V,2013A&A...549A.147B} by exploring low-latitude regions where studies of stellar ages were unobtainable. A hitherto unobserved component of the inner Galaxy was identified, namely a young inner thin disk along the Galactic mid-plane. The confined vertical extent, together with the wide longitudinal breadth of the Cepheids' spatial distribution, suggest that this stellar disk has a smooth transition from both the nuclear bulge and the Galactic thin disk that encompasses the bulge region.

The findings discussed here are in general agreement with recent numerical results concerning the ages of stellar constituents residing in the inner Galaxy. In simulated disk galaxies forming stars from cooling gas inside a dark matter halo, a sizable population of young and metal-rich stars emerged in close proximity to the plane, within a radius of $\sim 2$~kpc from the centers.  Having a median age of  $\sim1$~Gyr and a large age dispersion, these simulated stars show evidence that the presence of young stars in an old bulge can follow naturally from a model galaxy's evolution \citep{2014ApJ...787L..19N}. Yet continued investigations are needed to assess whether the Cepheids in the inner disk were born \emph{in situ}, possibly from star formation stemming from gas inflow towards the GC \citep[e.g.,][]{2004ARA&A..42..603K}, or if they originate from further out, either being formed at the bar ends \citep{1996ASPC...91...44P} or captured from the outer disk by a growing bar \citep{2006ApJ...637..214M}. Constraining the Cepheids' orbits by kinematical measurements would supply critical information for resolving this point. Another important question is whether the young stars in the inner disk evolve dynamically as old bulge stars once did, and thus will eventually flare into the boxy outer bulge. Such a secular process could supply younger metal-rich stars to the predominantly old outer bulge \citep{2013A&A...549A.147B}. Given that disk galaxies with small bulges such as ours are common, understanding their stellar components' fundamental properties, interactions, and coevolution is key in our quest to understand galaxy evolution as a whole.

\acknowledgments
We gratefully acknowledge the use of data from the ESO Public Survey program 179.B-2002, taken with the VISTA telescope, and data products from the Cambridge Astronomical Survey Unit. The authors acknowledge the following funding sources: BASAL CATA PFB-06, the Chilean Ministry of Economy's ICM grant IC120009, FIC-R Fund (project 30321072), CONICYT-PCHA (Doctorado Nacional 2014-63140099), CONICYT Anillo ACT 1101, and FONDECYT projects 1141141, 1130196, 3130552, 1120601, and 1150345.

{\it Facilities:} \facility{ESO:VISTA}.

\clearpage

\clearpage

\begin{deluxetable}{cccrcrcrcccccccc}
\tabletypesize{\tiny}
\rotate
\tablecaption{Catalog of the classical Cepheids.\label{table}}
\tablewidth{0pt}
\tablehead{
\colhead{Obj.} & \colhead{RA} & \colhead{Dec.} & \colhead{Period} & \colhead{$a_{\rm tot.}$} & 
\colhead{$d_{\rm T_1}$} &\colhead{$A_{K_s,{\rm T_1}}$} & \colhead{$d_{\rm T_2}$} & \colhead{$A_{K_s,{\rm T_2}}$} &
\colhead{$A_{K_s,{\rm B}}$} & \colhead{$\sigma A_{K_s,{\rm RC}}$} & \colhead{$K_s$} & \colhead{$H-K_s$} & \colhead{$J-K_s$} & \colhead{Note} \\
 & \colhead{[J2000.0]} & \colhead{[J2000.0]} & \colhead{[days]} & \colhead{[mag.]} & \colhead{[kpc]} & \colhead{[mag.]} & \colhead{[kpc]} & \colhead{[mag.]} &
\colhead{[mag.]} &\colhead{[mag.]} &\colhead{[mag.]} &\colhead{[mag.]} &\colhead{[mag.]} &
}
\startdata
1   & 17:22:23.24 & -36:21:41.5 &   7.5685 & 0.22 & 10.99  (0.96) & 2.07 & 4.40  (0.40) & 1.94 & 1.44 & 0.21 & 12.01 (.02)& 1.31  (.07) & 3.68 (.03) & d \\
2   & 17:21:16.05 & -36:43:25.2 &   6.3387 & 0.21 & 9.97    (1.02) & 2.59 & 4.11  (0.43) & 2.47 & 1.59 & 0.20 & 12.57 (.02)& 1.62  (.10) & 4.65 (.07) & \\
3   & 17:20:14.62 & -37:11:16.0 &   5.0999 & 0.20 & 9.55    (0.91) & 1.84 & 4.07  (0.40) & 1.72 & 1.21 & 0.22 & 12.04 (.02)& 1.16  (.08) & 3.21 (.04) & \\
4   & 17:26:34.72 & -35:16:24.1 & 13.4046 & 0.23 & 9.57    (0.77) & 2.88 & 3.49  (0.28) & 2.74 & 2.02 & 0.21 & 11.70 (.02)& 1.81  (.05) & 5.24 (.09) & \\
5   & 17:25:29.70 & -34:45:45.9 & 12.3267 & 0.29 & 10.23  (0.80) & 2.78$^*$ & 3.78  (0.30) & 2.64 & 1.79 & 0.12 & 11.86 (.02)& 1.75  (.04) & 5.01 (.07) & d \\
6   & 17:26:43.41 & -34:58:25.6 &   9.8383 & 0.22 & 9.99    (1.15) & 4.50$^*$ & 3.83  (0.44) & 4.37 & 1.77 & 0.40 & 13.86 (.02)& 2.80  (.11) & \dots & \\
7   & 17:26:54.24 & -35:01:08.2 &   4.2904 & 0.20 & 10.79  (0.95) & 3.23$^*$ & 4.74  (0.44) & 3.11 & 1.64 & 0.32 & 13.93 (.02)& 2.01  (.05) & \dots &  \\
8   & 17:30:46.64 & -34:09:04.4 &   4.7272 & 0.13 & 9.46    (0.84) & 3.13$^*$ & 4.09  (0.38) & 3.01 & 1.15 & 0.21 & 13.41 (.02)& 1.95  (.06) & \dots  &  \\
9   & 17:28:15.86 & -34:32:27.2 &   7.0235 & 0.22 & 8.50    (0.75) & 3.78$^*$ & 3.44  (0.31) & 3.65 & 1.67 & 0.15 & 13.27 (.02)& 2.35  (.06) & \dots  & e,[1]  \\
10 & 17:38:42.96 & -31:44:55.7 & 11.9663 & 0.28 & 10.56  (0.83) & 2.81 & 3.92  (0.31) & 2.67 & 1.61 & 0.22 & 12.01 (.03)& 1.77  (.04) & 4.38 (.09) & \\
11 & 17:36:44.46 & -32:04:38.6 &   8.3220 & 0.17 & 6.58    (0.65) & 4.40$^*$ & 2.59  (0.26) & 4.27 & 1.31 & 0.33 & 13.09 (.03)& 2.74  (.07) & \dots  & \\
12 & 17:40:41.72 & -30:48:46.9 & 10.6634 & 0.22 & 9.89    (0.90) & 2.91 & 3.74  (0.34) & 2.78 & 1.73 & 0.27 & 12.13 (.04)& 1.83  (.07) & 5.26 (.08) & \\
13 & 17:40:25.15 & -31:04:50.5 &   4.7004 & 0.16 & 7.92    (0.69) & 2.19 & 3.42  (0.31) & 2.07 & 1.79 & 0.22 & 12.10 (.04)& 1.38  (.06) & 3.94 (.05) & \\
14 & 17:42:20.00 & -30:14:50.7 & 23.9729 & 0.42 & 9.09    (0.91) & 4.16$^*$ & 3.02  (0.29) & 4.01 & 1.51 & 0.25 & 12.05 (.04)& 2.61  (.07) & \dots  & d \\
15 & 17:41:15.13 & -30:07:17.7 & 13.3262 & 0.33 & 9.31    (1.00) & 2.84$^*$ & 3.40  (0.36) & 2.70 & 1.22 & 0.19 & 11.62 (.04)& 1.79  (.10) & 5.26 (.14) & d \\
16 & 17:51:05.72 & -26:38:18.3 &   6.3496 & 0.17 & 9.62    (0.81) & 2.95 & 3.96  (0.34) & 2.82 & 1.77 & 0.34 & 12.85 (.03)& 1.84  (.05) & 5.26 (.07) & dey \\
17 & 17:51:13.77 & -26:48:55.9 & 12.9488 & 0.25 & 10.25  (0.90) & 3.11$^*$ & 3.76  (0.33) & 2.96 & 1.41 & 0.27 & 12.13 (.03)& 1.95  (.06) & \dots & e \\
18 & 17:49:41.42 & -27:27:14.6 & 10.4762 & 0.20 & 9.81    (0.81) & 2.25 & 3.72  (0.31) & 2.12 & 1.46 & 0.35 & 11.48 (.03)& 1.43  (.06) & 4.03 (.04) & \\
19 & 17:50:30.49 & -27:13:46.7 & 12.6433 & 0.29 & 8.28    (0.72) & 3.70$^*$ & 3.05  (0.26) & 3.56 & 1.59 & 0.55 & 12.30 (.03)& 2.32  (.05) & \dots & \\
20 & 17:53:16.07 & -26:28:26.9 &   5.9995 & 0.21 & 7.57    (0.65) & 3.62$^*$ & 3.15  (0.28) & 3.50 & 1.35 & 0.36 & 13.08 (.03)& 2.25  (.05) & \dots & \\
21 & 17:52:21.66 & -26:31:19.3 & 11.9921 & 0.27 & 9.73    (0.85) & 3.23$^*$ & 3.61  (0.31) & 3.09 & 1.69 & 0.29 & 12.25 (.03)& 2.02  (.06) & 5.29 (.08) & d \\
22 & 17:51:43.80 & -26:31:11.3 &   5.3407 & 0.15 & 7.78    (0.68) & 3.57$^*$ & 3.30  (0.30) & 3.45 & 1.47 & 0.27 & 13.25 (.03)& 2.22  (.05) & \dots & b,[2] \\
23 & 17:55:44.68 & -25:00:30.2 &   6.7911 & 0.20 & 9.95    (0.88) & 2.77$^*$ & 4.05  (0.37) & 2.64 & 1.38 & 0.32 & 12.64 (.03)& 1.73  (.07) & 5.01 (.10) & \\
24 & 17:58:26.68 & -23:52:08.3 &   4.5517 & 0.16 & 9.10    (0.87) & 3.12$^*$ & 3.96  (0.39) & 3.00 & 1.22 & 0.23 & 13.37 (.02)& 1.94  (.07) & 5.44 (.22) & \\
25 & 18:03:31.12 & -22:21:14.0 & 10.9622 & 0.16 & 8.70    (0.71) & 3.55$^*$ & 3.28  (0.27) & 3.41 & 1.92 & 0.13 & 12.45 (.02)& 2.22  (.04) & \dots & dey \\
26 & 18:09:14.03 & -20:03:21.4 & 24.8683 & 0.42 & 7.28    (0.72) & 5.64$^*$ & 2.40  (0.23) & 5.49 & 0.83 & 0.24 & 13.00 (.02)& 3.52  (.04) & \dots & de,[3] \\
27 & 17:22:10.10 & -36:44:18.8 & 14.8719 & 0.33 & 9.55    (0.95) & 2.77$^*$ & 3.43  (0.34) & 2.63 & 1.47 & 0.28 & 11.44 (.02)& 1.75  (.09) & 5.19 (.05) & b,[4] \\
28 & 17:26:00.10 & -35:15:15.0 & 18.2396 & 0.21 & 8.80    (0.71) & 2.88 & 3.05  (0.24) & 2.73 & 2.39 & 0.07 & 11.09 (.02)& 1.82  (.04) & 5.21 (.10) & \\
29 & 17:32:14.07 & -33:23:59.5 &   9.9061 & 0.21 & 9.86    (0.82) & 1.69 & 3.78  (0.32) & 1.56 & 1.41 & 0.19 & 11.01 (.02)& 1.08  (.07) & 2.98 (.06) & \\
30 & 17:40:51.51 & -30:24:53.2 &   7.9311 & 0.17 & 9.75    (0.94) & 2.11 & 3.87  (0.38) & 1.97 & 1.60 & 0.23 & 11.72 (.04)& 1.33  (.09) & 3.70 (.06) & \\
31 & 17:40:24.58 & -31:01:32.9 & 17.3711 & 0.32 & 10.57  (0.93) & 2.50 & 3.70  (0.32) & 2.36 & 1.94 & 0.29 & 11.18 (.04)& 1.59  (.06 )& 4.62 (.09) & \\
32 & 17:50:17.54 & -27:08:13.3 &   7.0503 & 0.14 & 10.22  (0.88) & 2.47 & 4.14  (0.37) & 2.34 & 1.62 & 0.52 & 12.35 (.03)& 1.55  (.06) & 4.44 (.06) & \\
33 & 17:55:24.20 & -25:30:22.3 & 15.4472 & 0.37 & 10.24  (0.88) & 2.69 & 3.65  (0.31) & 2.54 & 1.44 & 0.33 & 11.46 (.03)& 1.70  (.06) & 5.14 (.05) & \\
34 & 17:54:40.25 & -25:34:39.5 & 17.1620 & 0.41 & 8.82    (0.78) & 3.17$^*$ & 3.09  (0.27) & 3.02 & 1.42 & 0.30 & 11.47 (.03)& 2.00  (.06) & 5.86 (.09) & y,[5] \\
35 & 17:56:01.96 & -25:15:44.9 & 13.4540 & 0.25 & 9.78    (0.85) & 2.55$^*$ & 3.57  (0.31) & 2.41 & 1.14 & 0.20 & 11.42 (.03)& 1.61  (.07) & 4.59 (.04) & \\
\enddata
\tablecomments{
Errors are given in parentheses. $a_{\rm tot.}$: $K_s$ total amplitude; $d$: Heliocentric distance; $A_{K_s}$: $K_s$ extinction; $A_{K_s,B}$: extinction from \cite{2012A&A...543A..13G}; 
T1/T2: quantities under the assumptions that the object is classical or type II Cepheid, respectively.\\
Objects in the immediate vicinity of the Cepheids:
[1]: DOBASHI 7236 (dark nebula); 
[2]: [CWP2007] CN 35 (HII bubble); 
[3]: \cite{2010ApJ...719.1104R} SFC02 star forming complex; 
[4]: [CWP2007] CS 96 (HII bubble); 
[5]: SSTGLMC G003.9104+00.0010 (young stellar object)
}
\tablenotetext{*}{$A_{K_s,T1}$ has $>3\sigma$ positive deviation from $A_{K_s,{\rm B}}$.}
\tablenotetext{b}{Known ionozed hydrogen bubble close to the Cepheid.}
\tablenotetext{d}{Elevated diffuse mid-infrared emission around the Cepheid at 8- and 24-$\mu m$ GLIMPSE images.}
\tablenotetext{e}{Extended bright object close to the Cepheid in 8- and 24-$\mu m$ GLIMPSE images.}
\tablenotetext{y}{Young stellar object(s) close to the Cepheid.}

\end{deluxetable}

\end{document}